\documentclass[conference]{IEEEtran}

\usepackage{cite}

\ifCLASSINFOpdf
  \usepackage[pdftex]{graphicx}
  \graphicspath{{pdf/}{jpeg/}{media/}}
  \DeclareGraphicsExtensions{.pdf,.jpeg,.png}
\else
\fi

\usepackage{url}
\usepackage{textcomp}
\usepackage{booktabs}
\usepackage{tabularx}
\usepackage[utf8]{inputenc}
\usepackage[T1]{fontenc}
\usepackage{float}
\usepackage{footnote}
\usepackage{subfigure}
\usepackage{color}
\usepackage{listings}
\usepackage{amsmath}
\usepackage{tcolorbox}
\usepackage{etoolbox}
\usepackage{moresize}
\usepackage{fixltx2e}
\usepackage{enumitem}

\usepackage[frozencache=true,cachedir=.]{minted}

\usepackage{fancyhdr}
\fancypagestyle{firstpage}{%
  \rhead{\bfseries \textcolor{red}{Accepted for publication in 9th IEEE International Conference on Software Engineering and Service Science (ICSESS 2018) on 09.07.2018 - published version may differ}}
}

\begin{document}

\title{Split-Scale: Scaling Bitcoin by Partitioning the UTXO Space}

\author{
\IEEEauthorblockN{Kaz{\i}m R{\i}fat \"{O}zy{\i}lmaz}
\IEEEauthorblockA{Department of Computer Engineering\\
Bogazici University\\
Istanbul, Turkey\\
kazim@monolytic.com}
\and
\IEEEauthorblockN{Harsh Patel}
\IEEEauthorblockA{Ahmedabad, India\\
reach@harshpatel.space}
\and
\IEEEauthorblockN{Ankit Malik}
\IEEEauthorblockA{New Delhi, India\\
ankit@ankitmalik.in}
}

\maketitle

\thispagestyle{firstpage}

\begin{abstract}
The Bitcoin protocol is a significant milestone in the history of money. However, its adoption is currently constrained by the transaction limits of the system. As the chief problem of blockchain technology, the scaling issue has attracted many valuable solutions both on-chain and off-chain.

In this paper, our goal is to explore the notion of unspent transaction outputs (UTXOs) to propose an augmented Bitcoin protocol that can scale gracefully. Our proposal aims to increase the transaction throughput by partitioning the UTXO space and splitting the blockchain. In addition, a new type of Bitcoin node is introduced to preserve the capability to run validating nodes in low-bandwidth environments, despite the increased transaction throughput.
\end{abstract}


\IEEEpeerreviewmaketitle

\section{Introduction}
Scalability is the one of the most important aspects affecting Bitcoin\textquotesingle s adoption. Limits of scalability express themselves as high transaction fees, which affects usability and adoption negatively. To improve transaction throughput, various proposals have been made, starting with directly increasing block size. However, the most effective scaling improvement already integrated to Bitcoin is 'Segregated Witness'~\cite{segwit}. This increased the block capacity by introducing the block weight metric. Another notable attempt to solve the scalability problem is the Bitcoin-NG protocol~\cite{eyal2016bitcoin} which introduced an additional mining process where miners gain the capability to mine microblocks by mining a regular Bitcoin block.

In this paper, we propose a solution that increases the transaction throughput of the Bitcoin network without hurting network decentralization in terms of bandwidth requirements. By partitioning the UTXO space and splitting the blockchain into a tree structure, independently operating sub-chains will be created at every split event. As a result, a new block from all sub-chains will be mined at every block interval, increasing the transaction throughput exponentially. Moreover, in order to preserve the capability to run a node in this increasing bandwidth requirement, a new type of Bitcoin node (the half node) is introduced. Although this new node type does not store the complete blockchain, it can independently verify the transactions on the sub-chain it is tracking, which gives it the capability to operate in low-bandwidth environments.

In the next section (Section~\ref{sec_coreconcepts}), an overview of the core concepts is presented. Then, the general mechanics and technical details (Section~\ref{sec_splitscale}) of the proposal are described. Effects of the proposal to mining (Section~\ref{sec_mining}) and network organization (Section~\ref{sec_network}) are discussed in the following sections. Next, the transactions discussion (Section~\ref{sec_txs}) provides insights on transactions in the proposed system. A section dedicated to comparison of the split-scale proposal to other major Bitcoin scaling solutions (Section~\ref{sec_comparison}) is included afterwards.

\section{Core Concepts}
\label{sec_coreconcepts}

\subsection{Unspent Transaction Output (UTXO)}
\label{sec_utxo}
Bitcoin does not use the concept of 'account balance' as Ethereum does. Instead, total balance of a Bitcoin account is the accumulated amount of the \textit{transaction outputs} that are claimable but not yet spent. These \textit{unspent transaction outputs} or \textit{UTXOs} for short, are used as the inputs of the transactions. They are referred using the \textit{source transaction hash} and \textit{index of the output} within that source transaction (Listing~\ref{list:code}).

\begin{listing}
\begin{minted}[mathescape, linenos, numbersep=5pt, frame=lines, framesep=2mm, fontsize=\scriptsize]{csharp}
    // transaction input
    class CTxIn
    {
    public:
	COutPoint prevout;   // UTXO-to-spend
	CScript scriptSig;   // input script
	uint32_t nSequence;
	CScriptWitness scriptWitness;
    };
    // pointer to transaction output
    class COutPoint
    {
    public:
	uint256 hash;        // transaction hash
	uint32_t n;          // index of the output
    };
    // transaction output
    class CTxOut
    {
    public:
	CAmount nValue;       // amount of Bitcoin
	CScript scriptPubKey; // output script
    };
\end{minted}
\caption{Transaction Input~\cite{btctxin}, OutPointer~\cite{btctxoutpointer} and Output~\cite{btctxout}}
\label{list:code}
\end{listing}

The receiving party will have at least one transaction output after the transaction is validated and added to the blockchain. If a UTXO is used as a transaction input and the transaction is already a part of the blockchain, then it is considered spent and thus can not be used a second time as a transaction input.

The UTXO set is stored by nodes in a database called \textit{chainstate.db} outside the blockchain, which provides persistent key-value storage. As of Bitcoin 0.15.0, the chainstate database has been changed from a per-transaction model to a per-output model which added benefits like faster serialization, predictable memory usage and better caching~\cite{chainstate}. This change may provide a smooth transition for the chain splitting mechanism that is proposed in this paper.

The \textit{CTxIn} class (Listing~\ref{list:code}, line 2-9) is a simplified version of the transaction input. It contains the location of the previous transaction\textquotesingle s output that it claims and a signature that matches the output's public key. The \textit{COutPoint} class (Listing~\ref{list:code}, line 11-16) in the transaction input shows how UTXOs are actually referred by the input. It contains both the \textit{transaction hash} and the \textit{index} of its output. Lastly, the \textit{CTxOut} class (Listing~\ref{list:code}, line 18-23) presents the anatomy of a transaction output in a simplified form. It contains the amount and the script \textit{scriptPubKey} to claim the output. Below is the \textit{scriptPubKey} for a standard Pay-to-PubkeyHash (P2PKH) transaction~\cite{btcscript}:

\begin{tcolorbox}[fontupper=\scriptsize] 
OP\_DUP OP\_HASH160 pubKeyHash OP\_EQUALVERIFY OP\_CHECKSIG
\end{tcolorbox}

On a side note, as of 26th of October 2017, 82\% of the Bitcoin transactions are Pay-to-PubkeyHash (P2PKH)~\cite{delgadoanalysis} so \textit{scriptPubKey} of these transactions are directly tied to a single receiving address.

\subsection{Memory Pool (mempool)}
\label{sec_mempool}
The \textit{memory pool}, or \textit{mempool}, is the memory area reserved by Bitcoin clients to store unconfirmed transactions. Unconfirmed transactions are accumulated in the mempool until they are picked by a miner, mined, and added to the blockchain. Currently, each node maintains its own mempool, having the complete view of all unconfirmed transactions in the Bitcoin network. The amount of memory reserved for mempool varies greatly, the peak point being around 140MB for the last two years~\cite{mempool}.

\section{Split-Scale}
\label{sec_splitscale}
\begin{figure}[ht]
  \centering
  \includegraphics[scale=0.6]{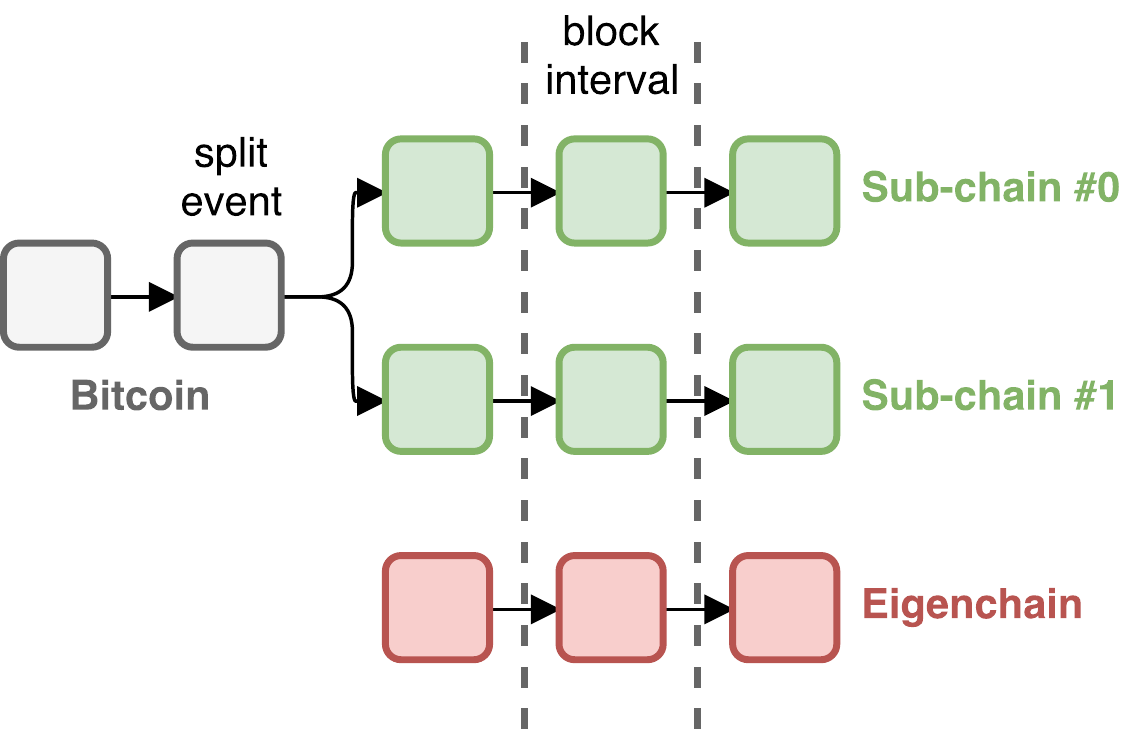}
  \caption{Bitcoin after Split Event}
  \label{fig:sequence}
\end{figure}

The idea is to split the Bitcoin blockchain (Figure~\ref{fig:sequence}), known here as \textit{split events}, into multiple sub-chains in order to:
\begin{itemize}[nosep]
  \item create independently operating multiple sub-chains, therefore creating multiple blocks instead of one block for every block creation interval (an interval lasts 10 minutes on average).
  \item provide the flexibility of operating home nodes with less bandwidth and storage requirements. These nodes will have the option to track only a subset of chains without losing any verification capability.
\end{itemize}
The mechanics of such a split and how UTXO database, mempool or mining operations will be affected will be presented in the following subsections.

\subsection{Split Event}
\label{sec_splitevent}
A \textit{split event} is a deterministic (and repeatable) action that will be triggered as a result of a decision made by the governing authority of the platform and/or certain performance metrics showing that the system is pushing its boundaries in terms of transaction throughput. Details on when a split event will occur or who is going to decide for it is a separate topic that will not be addressed in the scope of this paper. At the split event, following changes will happen on Bitcoin clients:
\begin{itemize}[nosep]
  \item UTXO database will be divided based on their \textit{scriptPubKey} hashes.
  \item the UTXO split may be implemented logically (UTXO hashes in binary form that start with 0 or 1 for the first split) or economically (find the 256bit number that divides the Bitcoin supply in half).
  \item the mempool will be divided based on which sub-chain they are tracking.
  \item miners must create a block for each sub-chain, and a separate block containing these block headers to claim a block reward.
\end{itemize}

\subsection{Dividing the Chainstate Database}
\label{sec_divutxo}
Based on the information given on the UTXO structure, it is possible to create a 256-bit hash value from the \textit{scriptPubKey} of any given UTXO. The hash function is selected as double SHA256() and a hash value representing all UTXOs with the same \textit{scriptPubKey} can be given as follows:

\begin{tcolorbox}[fontupper=\small] 
hash\textsubscript{UTXO} = SHA256(SHA256(CTxOut.scriptPubKey));
\end{tcolorbox}

This approach will work for all the script types including Pay-to-PubkeyHash (P2PKH), Pay-to-ScriptHash (P2SH), Pay-to-multisig (P2MS) or Pay-to-Pubkey (P2PK) outputs.

At the split event, the Bitcoin client will calculate 256-bit hash values for all UTXOs and decide in which chainstate database a UTXO will end up. At that point, the Bitcoin client will create \textit{chainstate0} and \textit{chainstate1} databases and remove the original \textit{chainstate.db}, in case of a two-way split.

From the split event onward, sub-chains start to add their own blocks. After the first valid blocks are added, two sub-chains behave just like mini-Bitcoin networks independent of each other.

\subsection{Dividing Mempool}
\label{sec_divmempool}
At the split event, Bitcoin memory pool will flush, create \textit{mempool0} and \textit{mempool1} and remove the original mempool. Starting with the next block, only transactions that belong to a specific sub-chain will be considered valid. Each sub-chain will transmit its unconfirmed transactions to a different mempool, and they will be picked up by miners separately for each sub-chain.

\subsection{After the Split}
\label{sec_divblockchain}
Dividing a UTXO set and mempool will create mini-Bitcoins effectively (Figure~\ref{fig:sequence}). These mini-Bitcoins will function independently by conforming the rules below:
\begin{itemize}[nosep]
  \item every full node can divide UTXO set and mempool, and keep track of the divided sets independently based on the guidelines (as additional consensus parameters coded in the Bitcoin client).
  \item mempools will detect (using the new consensus parameters) and won't accept mixed transactions, e.g. users can't mix up UTXOs belonging to different sub-chains in a single transaction.
  \item the UTXO set will be tracked in multiple databases, on a per-subchain basis.
\end{itemize}
A side effect of this is the continued presence of regular Bitcoin addresses in all sub-chains. However, there may be a different number of UTXOs attached to those accounts on every sub-chain.
The effects of such an event on mining and network organization will be elaborated on in the following sections.

\section{Mining}
\label{sec_mining}
Bitcoin uses Proof-of-Work (PoW) consensus that utilizes double SHA256() as its hash function. Every new block should contain the hash of the previous block in its header and the checksum of its own header should be lower than the 256-bit difficulty value that is updated every week. The proposed scaling solution will not attempt to change the hash function or propose a new consensus function. The aim of this paper is to adapt the current mining approach to a multi-chain setup.

As described in previous section, after the split event there are multiple sub-chains that can operate independently. This means every sub-chain may separately mine their own blocks and append to the blockchain, assuming all mining parties are honest. However, to keep the Bitcoin system robust and trustless in a multi-chain setup, some form of super-mining should be enforced, instead of making assumptions about the honesty of the other parties. Otherwise, miners with great computing power will jump through sub-chains depending on the difficulty values, which will make the system more vulnerable to instantaneous attacks~\cite{eyal2014majority}.

\subsection{Eigenchain}
In order to keep to mining power in check, adding a new block should happen atomically across on all sub-chains. This means the block count will be the same across all sub-chains, all the time. However, to verify the newly added blocks and detect double-spend attacks, there should be a separate blockchain that keeps track of the all the blocks added to their respective sub-chains. This new blockchain that stores the block header hashes of sub-chain blocks is called \textit{eigenchain}.

The mining will works as follows:
\begin{enumerate}
\item Miners have to listen to all sub-chains and pick transactions from all mempools.
\item Miners have to mine a block for every sub-chain (the difficulty levels will be adjusted after the split events).
\item Miners have to create a new eigenchain block by using sub-chain block headers. The difficulty of the eigenchain block should be higher than the sub-chain blocks, almost having the same difficulty in Bitcoin network at the time of the split event.
\end{enumerate}

In Bitcoin, a newly mined block is serialized~\cite{btcserializedblocks} and transmitted using the \textit{block} message~\cite{btcblock}. In this proposal, however, the \textit{block} message will transmit a single serialized block similar to Bitcoin but that block will contain both the new eigenchain block and all the other sub-chain blocks.

Although the proposed approach seems like a glorified block size increase at the moment, the changes in the network organization and introduction of the \textit{half node} will show the benefits of the approach. For an explicit comparison, refer to the Section~\ref{sec_comparison}: "Bitcoin Proposal Comparison." The network organization will be discussed in the following section.

\begin{table*}[htbp]
\centering
\caption{Split-Scale Comparison: Miner Perspective}
\label{tab:miner}
\begin{tabular}{lllll}
\toprule
Miner Features &
\begin{tabular}[t]{@{}l@{}}Bitcoin (SegWit)\end{tabular} &
\begin{tabular}[t]{@{}l@{}}SegWit2x\end{tabular} &
\begin{tabular}[t]{@{}l@{}}Bitcoin-NG\end{tabular} &
\begin{tabular}[t]{@{}l@{}}Split-Scale\end{tabular}\\
\midrule
Scale Factor &
\begin{tabular}[t]{@{}l@{}}1x\end{tabular} &
\begin{tabular}[t]{@{}l@{}}2x\end{tabular} &
\begin{tabular}[t]{@{}l@{}}60x\end{tabular} &
\begin{tabular}[t]{@{}l@{}}Nx\\(scales exponentially with split count)\end{tabular} \\
\midrule
Block Count &
\begin{tabular}[t]{@{}l@{}}mine one block\end{tabular} &
\begin{tabular}[t]{@{}l@{}}mine one block\end{tabular} &
\begin{tabular}[t]{@{}l@{}}mine one key block\\plus microblocks (every 10s)\end{tabular} &
\begin{tabular}[t]{@{}l@{}}mine one block on all sub-chains\\plus one eigenchain block\end{tabular} \\
\midrule
Block Size &
\begin{tabular}[t]{@{}l@{}}\textasciitilde 1MB on average\\4MB SegWit limit\end{tabular} &
\begin{tabular}[t]{@{}l@{}}\textasciitilde 2MB on average\\8MB SegWit limit\end{tabular} &
\begin{tabular}[t]{@{}l@{}}same as Bitcoin (SegWit)\end{tabular} &
\begin{tabular}[t]{@{}l@{}}same as Bitcoin (SegWit)\end{tabular} \\
\midrule
Transaction Fees &
\begin{tabular}[t]{@{}l@{}}from one block\end{tabular} &
\begin{tabular}[t]{@{}l@{}}from one block\end{tabular} &
\begin{tabular}[t]{@{}l@{}}from all key and microblocks\end{tabular} &
\begin{tabular}[t]{@{}l@{}}from all sub-chain blocks\end{tabular} \\
\bottomrule
\end{tabular}
\end{table*}

\begin{table*}[htbp]
\centering
\caption{Split-Scale Comparison: Node Perspective}
\label{tab:node}
\begin{tabular}{lllll}
\toprule
Node Requirements &
\begin{tabular}[t]{@{}l@{}}Bitcoin (SegWit)\end{tabular} &
\begin{tabular}[t]{@{}l@{}}SegWit2x\end{tabular} &
\begin{tabular}[t]{@{}l@{}}Bitcoin-NG\end{tabular} &
\begin{tabular}[t]{@{}l@{}}Split-Scale\end{tabular}\\
\midrule
Storage Requirements &
\begin{tabular}[t]{@{}l@{}}whole blockchain\end{tabular} &
\begin{tabular}[t]{@{}l@{}}whole blockchain\end{tabular} &
\begin{tabular}[t]{@{}l@{}}whole blockchain\end{tabular} &
\begin{tabular}[t]{@{}l@{}}full nodes store the whole blockchain\\half nodes store only one sub-chain\\and eigenchain
\end{tabular} \\
\midrule
Bandwidth Requirements &
\begin{tabular}[t]{@{}l@{}}at least \textasciitilde 700Kb~\cite{btcminreq}\end{tabular} &
\begin{tabular}[t]{@{}l@{}}at least \textasciitilde 1.4Mb\end{tabular} &
\begin{tabular}[t]{@{}l@{}}full node bandwidth requirements\\increase linearly with scaling factor\\(60x)\end{tabular} &
\begin{tabular}[t]{@{}l@{}}full node bandwidth requirements\\increase linearly with scaling factor (Nx)\\half node requirements will be similar\\to Bitcoin (SegWit)\end{tabular} \\
\bottomrule
\end{tabular}
\end{table*}

\section{Network}
\label{sec_network}
The Bitcoin network consists of multiple types of peers: miners, full nodes and lightweight nodes. Miners are the peers that create and transmit new blocks to the network, full nodes are the verifiers that store the complete blockchain and lightweight nodes are the relatively weak ones that use \textit{Simple Payment Verification (SPV)} to only verify particular transactions~\cite{btcspv}.

\subsection{Full Nodes}
In the regular Bitcoin network, full nodes store the complete blockchain and execute block and transaction verifications all the time to keep the system secure. Similarly, in our proposal, full nodes will keep in sync with all sub-chains plus the eigenchain, therefore it will be able to verify a specific sub-chain in itself and cross-reference it with the eigenchain. Miners and full nodes are connected in a way similar to the current Bitcoin network formation and new \textit{block} messages are only sent to full nodes. Full nodes will verify and update the newly mined blocks, and will then re-transmit the sub-chain blocks (a serialized eigenchain block and appended sub-chain block) to the relevant networks formed by sub-chain nodes. In short, full nodes are interconnected to full nodes and half nodes. Not all messages are sent to sub-chain networks, however. Only the relevant ones are propagated to minimize the bandwidth requirements.

\subsection{Half Nodes}
With the proposed scheme, an additional type of node called \textit{half node} is added to the system. Half nodes keep track of one sub-chain and the eigenchain. A half node is able to verify both the tracked sub-chain and eigenchain blocks by using block header hashes and is able to cross-check and validate sub-chain blocks using information contained in eigenchain. Half nodes only keep track of one mempool and one chainstate database (UTXO set) depending on which sub-chain they select. In addition, half nodes do not get new \textit{block} messages. New blocks targeting sub-chains are transmitted by using a new type of message: a \textit{block-n} message, which contains only the serialized eigenchain and sub-chain block. This way both the storage and bandwidth requirements of half node will be significantly lower compared to full nodes.

\section{Transactions}
\label{sec_txs}
After the split event, all the UTXOs of a specific \textit{scriptPubKey} will be accumulated in one sub-chain. Basically, users will be able to create transactions using only UTXOs from a specific sub-chain and be able to transact without knowing the remaining sub-chains. However, in time, users will receive payments from multiple parties in different sub-chains. Therefore total account balance of a user will more or less reside in multiple sub-chains with different UTXOs attached to it. If a user wants to spend more than the total amount of Bitcoin in one of his sub-chains then multiple transactions should be made.

\subsection{Hashed Time-Lock Contract (HTLC)}
\textit{Hashed Time-Lock Contract}, or \textit{HTLC} in short, is defined as: 'a class of payments that use hashlocks and timelocks to require that the receiver of a payment either acknowledge receiving the payment prior to a deadline by generating cryptographic proof of payment or forfeit the ability to claim the payment, returning it to the payer'~\cite{btchtlc}. Lightning Networks use HTLC to be able to construct secure transfers using a network of channels across multiple hops to the final destination~\cite{poon2016bitcoin}.

In the proposed system, HTLCs are used to ensure the atomicity of the payment, even the payment consists of multiple transactions on multiple sub-chains. Assuming that the sender does not have enough balance on one sub-chain to cover the complete payment, then the sender has to create multiple transactions on multiple sub-chains respectively. The receiver may claim each transaction on a different sub-chain, but it is preferred to finalize the payment in a single step. In such cases, senders (therefore the underlying Bitcoin wallet implementations) should utilize HTLCs to ensure atomicity. The sending process should be as follows:
\begin{enumerate}
\item the sender creates random data.
\item the hash of that random data is calculated.
\item the hash value is added to all transactions (\textit{scriptPubKey}) and transactions are sent on their respective sub-chains.
\item after all the transactions are mined, the total payment amount may be claimed by the receiver, after complete random data is shared by the sender.
\item if any of the transactions fails in a predefined time interval, funds may be claimed by the sender again.
\end{enumerate}

\subsection{Eigentransactions}
\label{sec_eigentxs}
An eigentransaction is a failsafe mechanism which may be added to the system to make fund transfer possible between the sub-chains. However, these transactions are special and limited to sending funds only between the same addresses in multiple sub-chains, so the private key of sending and claiming address should be the same. This is to provide an easy way for transferring the total account balance into a single sub-chain.

Eigentransactions should have a separate global pool called the \textit{eigenpool} similar to the current Bitcoin mempool, and eigentransactions are mined and included into the eigenchain. This way all sub-chains will be able to track fund transfers of the same account across sub-chains and will be able to add the UTXO (if sent to that specific sub-chain) to their balance.

With the addition of eigentransactions, the block size of the eigenchain will be increased. However, activation of this feature can be easily controlled, even enabled/disabled between certain block numbers.

\section{Bitcoin Proposal Comparison}
\label{sec_comparison}
Two forms of decentralization are at the heart of the Bitcoin scaling debate. The first form is mining decentralization, which is the problem of accumulation of high hash rates at the hands of a limited number of mining cartels. The second one is decentralization of the network, which is the decreasing amount of full nodes due the increasing bandwidth requirements. Our proposal aims to scale the Bitcoin network without decreasing network decentralization. In Table~\ref{tab:miner} and Table~\ref{tab:node}, the split-scale proposal is compared to the other valuable on-chain scaling proposals in terms of miners and network node features.

Split-scale provides a framework for scaling and gives the opportunity to scale exponentially with every split event. For the miner, our proposal will provide better economic incentives, because although the block reward is the same, the transaction fees will be collected from all sub-chain blocks. As a result, transaction fee gains for miners will even surpass Bitcoin-NG at the sixth split event (64 sub-chains) (Table~\ref{tab:miner}).

Finally, our solution is clearly efficient in terms of bandwidth and storage requirements (Table~\ref{tab:node}). In all the other proposals transaction throughput increase (scaling) is directly translated to bandwidth and storage increase for Bitcoin nodes. As a result, due to the increasing requirements, the number of Bitcoin full nodes will decrease and network decentralization will suffer in all the other proposals. Split-scale introduces a new kind of Bitcoin node which is called 'half-node' that eliminates these restrictions and provides capability to run a node tracking only one sub-chain and eigenchain.

\section{Conclusion}
Scalability is important for expanding adoption of Bitcoin. In this paper, we address the scalability problem by partitioning the UTXO space, therefore splitting the Bitcoin blockchain into multiple sub-chains. Our approach facilitates a block creation increase due to the mining taking place on all sub-chains and it proposes a way to still maintain nodes operating in low-bandwidth conditions. Compared with prominent Bitcoin scaling proposals, "split-scale" offers scalability while preserving network decentralization.

\bibliographystyle{IEEEtran}
\bibliography{IEEEabrv,icsess-bitcoin}

\end{document}